\documentclass
[prc,superscriptaddress,twocolumn,showpacs,preprintnu
mbers,amsmath,amssymb]{revtex4-1}
\usepackage[dvips]{graphicx}
\usepackage{dcolumn}
\usepackage{amsmath}
\usepackage{times}
\usepackage{braket}
\def\la{\mathrel{\mathpalette\fun <}}

\def\fun#1#2{\lower3.6pt\vbox{\baselineskip0pt\lineskip.9pt
 \ialign{$\mathsurround=0pt#1\hfil##\hfil$\crcr#2\crcr\sim\crcr}}}
\usepackage{graphicx}
\usepackage{dcolumn}
\usepackage{bm}

\newcommand{\beq}{\begin{equation}}
\newcommand{\eeq}{\end{equation}}
\newcommand{\bea}{\begin{eqnarray}}
\newcommand{\eea}{\end{eqnarray}}

\usepackage{ulem}
\usepackage{color}

\begin{document}


\title{Breakup and finite-range effects on the
$^{8}$B($d$,$n$)$^{9}$C reaction
}


\author{Tokuro Fukui}
\email[Electronic address: ]{tokuro@rcnp.osaka-u.ac.jp}
\author{Kazuyuki Ogata}%
\affiliation{%
 Research Center for Nuclear Physics, Osaka University, Osaka 567-0047, Japan
}%

\author{Masanobu Yahiro}
\affiliation{
 Department of Physics, Kyushu University, Fukuoka 812-8581, Japan
}%

\date{\today}

\begin{abstract}
The astrophysical factor of
$^8$B($p$,$\gamma$)$^9$C at zero energy, $S_{18}(0)$, is determined
by a three-body coupled-channels analysis of the transfer reaction
$^{8}$B($d$,$n$)$^{9}$C at 14.4~MeV/nucleon.
Effects of the breakup channels of $d$ and $^9$C are investigated with
the continuum-discretized coupled-channels method.
It is found that, in the initial and final channels, respectively,
the transfer process through the breakup states of
$d$ and $^9$C, its interference with that through
their ground states in particular,
gives a large increase in the transfer cross section.
The finite-range effects with respect to the proton-neutron relative
coordinate are found to be about 20\%.
As a result of the present analysis, $S_{18}(0)=22 \pm 6~{\rm eV~b}$ is
obtained, which is smaller than the result of the previous
distorted-wave Born approximation analysis by about 51\%.
\end{abstract}

\maketitle


\section{Introduction}
The explosive hydrogen burning called the hot $pp$ chain~\cite{Wiescher}
in low-metallicity supermassive stars
plays an important role as a possible alternative path
to the synthesis of the CNO elements.
The proton capture reaction of $^8$B, $^8$B($p$,$\gamma$)$^9$C, is
expected to lead to this hot $pp$ chain.
Since it is very difficult to  measure the cross section $\sigma_{p\gamma}$
for the $^8$B($p$,$\gamma$)$^9$C reaction at stellar energies,
several experiments of alternative reactions such as the
inclusive~\cite{Trache} and exclusive~\cite{Motobayashi} $^9$C breakup
reactions and the proton transfer reaction
$^{8}$B($d$,$n$)$^{9}$C~\cite{Beaumel} have been done to
determine the astrophysical factor 
\beq
S_{18}(\varepsilon_{p \rm B})=\sigma_{p\gamma} \varepsilon_{p \rm B} \exp[2\pi\eta].
\eeq
Here, $\varepsilon_{p \rm B}$ is the relative energy of the $p$-$^8$B
system in the center-of-mass (c.m.) frame and $\eta$ is the Sommerfeld parameter.
Because of the weak $\varepsilon_{p \rm B}$ dependence of
$S_{18}(\varepsilon_{p \rm B})$, its value at zero energy, $S_{18}(0)$, is
paid special attention as a reference value.

A problem with the results of the indirect measurements of $S_{18}(0)$
is that they are not consistent with each other, with values of 
$46\pm 6$~eV~b (from inclusive $^9$C breakup~\cite{Trache}),
$77\pm 15$~eV~b (from exclusive $^9$C breakup~\cite{Motobayashi}),
and $45\pm 13$~eV~b (from transfer~\cite{Beaumel}).
In Ref.~\cite{breakupS18}, reanalysis of the two $^9$C breakup reactions
has been performed with a three-body coupled-channels reaction model, and
$S_{18}(0)=66 \pm 10~{\rm eV~b}$ was obtained, resolving the
discrepancy between the two results of $^9$C breakup. There remains,
however, about a 30\% difference between the result of Ref.~\cite{breakupS18}
and that of the transfer reaction.
It was reported in Ref.~\cite{transferS17} that, in the
$^7$Be($d$,$n$)$^8$B reaction at 7.5~MeV, breakup channels of $d$
played an essential role. One may expect a similar effect also in
the $^{8}$B($d$,$n$)$^{9}$C reaction.

The purpose of the present study is to investigate the deuteron
breakup effects on the cross section of $^{8}$B($d$,$n$)$^{9}$C
at 14.4~MeV/nucleon and $S_{18}(0)$, by means of
the continuum-discretized
coupled-channels method (CDCC)~\cite{CDCC1,CDCC2,CDCC3}.
In the CDCC method, one non perturbatively treats the channel couplings of the
breakup (continuum) states of weakly bound nuclei,
and the method has been highly
successful in describing various real or virtual breakup reactions
in a wide range of incident energies. The theoretical foundation of
the CDCC method is given in Refs.~\cite{CDCC3,Aus89,Aus96}.
As an advantage over the previous CDCC study on
$^7$Be($d$,$n$)$^8$B~\cite{transferS17},
in this work the breakup channels of both the
{\lq\lq}projectile'' $d$, the target nucleus in inverse kinematics,
and the residual nucleus $^9$C are taken into account.
Furthermore,
a finite-range (FR) calculation of the transition matrix ($T$ matrix)
of the transfer reaction is performed.
We also propose a finite-range
correction (FRC) to the zero-range (ZR) calculation, which is appropriate
for three-body model calculation including breakup channels
of both the projectile and the residual nucleus.
Interpretation of the FR effects on $S_{18}(0)$ is given through this correction.

This paper is constructed as follows.
In Sec.~\ref{formulation}, we give a formulation of the coupled-channels
Born approximation (CCBA) for the $^{8}$B($d$,$n$)$^{9}$C reaction.
In Sec.~\ref{result}, we extract $S_{18}(0)$ from the transfer cross
section; the role of the breakup channels of $d$ and $^9$C are discussed.
The formalism of the FRC for the three-body reaction model and
discussion of the FR effects on the transfer cross section are also given.
Finally, we summarize this study in Sec.~\ref{summary}.

\section{Coupled-channels Born approximation (CCBA) formalism}
\label{formulation}
%
\begin{figure}[b]
\begin{center}
\includegraphics[width=0.4\textwidth,clip]{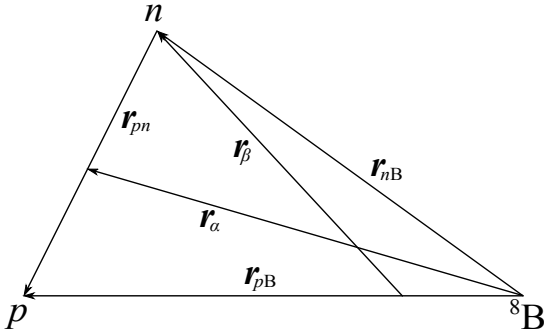}
\caption{Illustration of the three-body system.}
\label{fig1}
\end{center}
\end{figure}
In the present study we describe the transfer reaction
$^{8}$B($d$,$n$)$^{9}$C at 14.4~MeV/nucleon with the three-body
($p+n+{^8{\rm B}}$) model shown in Fig.~\ref{fig1}.
The transition matrix in the post form is given by
\begin{align}
T_{\beta \alpha}
&=
\Braket{
\Psi_{\beta}^{(-)}
|
V_{pn}
|
\Psi_{\alpha}^{(+)}
}
,\label{Tmat1}
\end{align}
where $\Psi_{\alpha}^{(+)}$ and $\Psi_{\beta}^{(-)}$ are, respectively,
three-body wave functions for the initial and final channels;
their explicit definition is given below.
The interaction between $p$ and $n$, $V_{pn}$, is adopted as the
transition interaction that causes the transfer process.
The superscripts ($+$) and ($-$) represent the outgoing and incoming
boundary conditions for the scattering wave, respectively.

The Schr\"odinger equation for $\Psi_{\alpha}^{(+)}$ is given by
\beq
\left[H_\alpha - E
\right]
\Psi_{\alpha}^{(+)}(\boldsymbol{r}_{pn},\boldsymbol{r}_{\alpha})
=0
,\label{Scheqi}
\eeq
\beq
H_\alpha=
K_{\boldsymbol{r}_{\alpha}} +  h_{pn}
 +U_{p \rm B}^{(\alpha)} ({r}_{p \rm B})
+U_{n \rm B}^{(\alpha)} ({r}_{n \rm B})
+V_{\rm C}(r_\alpha)
,\label{Hi}
\eeq
where $K_{\boldsymbol{X}}$
 is the kinetic energy operator
with respect to the coordinate $\boldsymbol{X}$,
$h_{pn}$ is the internal Hamiltonian of $d$,
and $E$ is the total energy of the three-body system.
The nuclear interaction between $x$ ($=p$ or $n$) and $^8$B is represented
by $U_{x \rm B}$ with the superscript $(\alpha)$ specifying the initial
channel. The Coulomb interaction between $d$ and $^8$B is denoted by
$V_{\rm C}$; we disregard the Coulomb breakup in this study.
We describe $\Psi_{\alpha}^{(+)}$ with CDCC as
\begin{align}
\Psi_{\alpha}^{(+)}(\boldsymbol{r}_{pn},\boldsymbol{r}_{\alpha})
&\approx
\sum_i
\psi_{pn}^i(\boldsymbol{r}_{pn})
\chi_\alpha^{ii_0(+)}(\boldsymbol{r}_{\alpha})
,\label{CDCCi}
\end{align}
where $\psi_{pn}^i$ is the internal wave function of $d$
with $i$ its energy index; $i=i_0$ corresponds to the ground state
of $d$ and $i\neq i_0$ to the discretized continuum states of
the $p$-$n$ system.
$\psi_{pn}^i$ satisfies
\begin{align}
\left(
h_{pn}-\varepsilon_{pn}^i
\right)
\psi_{pn}^i (\boldsymbol{r}_{pn})
=
0
,\label{CDwf}
\end{align}
where $\varepsilon_{pn}^i$ is the energy eigenvalue of the $p$-$n$ system.
One may obtain the $d$-$^8$B distorted wave $\chi_\alpha^{ii_{0}(+)}$
by solving the CDCC equations under the standard boundary
condition~\cite{CDCC1,CDCC2,CDCC3}.
Note that, in the present study, we ignore the intrinsic spin of each
particle for simplicity.
Details of the description of $\Psi_{\alpha}^{(+)}$ with CDCC
are given in Ref.~\cite{sbat}.

In the exact form of Eq. \eqref{Tmat1}, $\Psi_{\alpha}^{(+)}$ includes
not only the deuteron components, consisting of the elastic and
breakup ones, but also rearrangement components.
The latter are not explicitly taken into account
in the present CCBA calculation, which has been justified in Refs.~\cite{Aus89,Aus96}.

The three-body wave function $\Psi_{\beta}^{(+)}$ in the final channel,
which is the time reversal of $\Psi_{\beta}^{(-)}$, satisfies the
following Schr\"odinger equation:
\beq
\left[H_\beta - E
\right]
\Psi_{\beta}^{(+)}(\boldsymbol{r}_{p{\rm B}},\boldsymbol{r}_{\beta})
=0
,\label{Scheqf}
\eeq
\beq
H_\beta=
K_{\boldsymbol{r}_{\beta}} + h_{p{\rm B}}
+U_{n \rm B}^{(\beta)} ({r}_{n \rm B})
,\label{Hf}
\eeq
where $h_{p{\rm B}}$ is the $p$-$^8$B internal Hamiltonian given by
\beq
h_{p{\rm B}}=
K_{\boldsymbol{r}_{p \rm B}} +U_{p \rm B}^{(\beta)}(\boldsymbol{r}_{p \rm B})
+V_{\rm C}(r_{p \rm B}).
\eeq
The superscript $(\beta)$ represents the final channel.
Note that $H_\beta$ does not contain the term $V_{pn}$
that has been used as a transition interaction in Eq.~(\ref{Tmat1}).
In the CDCC framework
$\Psi_{\beta}^{(+)}$ is expressed by
\begin{align}
\Psi_{\beta}^{(+)}(\boldsymbol{r}_{p\rm B},\boldsymbol{r}_{\beta})
&\approx
\sum_j
\psi_{p \rm B}^j(\boldsymbol{r}_{p \rm B})
\chi_\beta^{jj_0(+)}(\boldsymbol{r}_{\beta})
,\label{CDCCf}
\end{align}
where
\beq
\left(
h_{p \rm B} -\varepsilon_{p{\rm B}}^j
\right)
\psi_{p \rm B}^{j}(\boldsymbol{r}_{p \rm B})
=0
\label{ScheqB}
\eeq
with $\psi_{p \rm B}^{j}$ the overlap functions of the ground and
discretized continuum states of $^9$C with the $p$-$^8$B(g.s.) configuration;
here the ground state is denoted by $j=j_0$ and $\varepsilon_{p{\rm B}}^j$ is
the eigenenergy of $^9$C in the $j$th state.
The $n$-$^9$C distorted wave
$\chi_{\beta}^{jj_0(+)}$ can be calculated with the same procedure as for $\chi_{\alpha}^{ii_0(+)}$.
Since the ground state of $^9$C includes the component that
cannot be described by the $p$-$^8$B(g.s.) configuration,
$\psi_{p \rm B}^{j_0}$ has to be normalized
by the square root of the spectroscopic factor $\cal S$.
The breakup components $\psi_{p \rm B}^{j}$ ($j \neq j_0$) also have to be
normalized by the same factor $\sqrt{\cal S}$, because
\bea
\Psi_\beta^{(+)} (\boldsymbol{r}_{p\rm B},\boldsymbol{r}_{\beta})
&=&\lim_{\epsilon \to +0}
\frac{i \epsilon}{E-H_{\beta}+i \epsilon}
e^{i \boldsymbol{k}_\beta \cdot \boldsymbol{r}_\beta} \sqrt{\cal S}
\psi_{p \rm B}^{j_0} (\boldsymbol{r}_{p\rm B})
\nonumber \\
&=& \sqrt{\cal S} \lim_{\epsilon \to +0}
\frac{i \epsilon}{E-H_{\beta}+i \epsilon}
e^{i \boldsymbol{k}_\beta \cdot \boldsymbol{r}_\beta}
\psi_{p \rm B}^{j_0} (\boldsymbol{r}_{p\rm B})
;
\nonumber \\
\label{Moller-1}
\eea
note that the $\psi_{p \rm B}^{j}$ ($j \neq j_0$) are generated by
the M$\phi$ller wave operator
${i \epsilon}/(E-H_{\beta}+i \epsilon)$.
Here, ${\cal S}$ has only one quantum number, i.e., $\ell=1$ for the
orbital angular momentum between $p$ and $^8$B(g.s.) in the ground state
of $^9$C. This is due to the neglect of the intrinsic spin of
each particle in the present study. Thus ${\cal S}$
is understood as 
an averaged value of the ${\cal S}$'s, each with a different value of the total angular
momentum
of the $p$-$^8$B(g.s.) system.


\section{Results and discussion}
\label{result}
\subsection{Model setting}
\label{result1}
We adopt the one-range Gaussian interaction~\cite{Ohmura} for
$V_{ pn}$. The pseudostate method
with the real-range Gaussian basis functions~\cite{Mat03}
is used for obtaining the discretized-continuum states of $d$;
we include the $s$ and $d$ states and neglect the intrinsic
spin of $d$. The number of basis functions taken is 20, and the
minimum (maximum) range parameter of the Gaussian is 1.0 (30.0)~fm.
We include in the CDCC pseudostates with $\varepsilon_{pn}^i < 65$~MeV and
$\varepsilon_{pn}^i < 80$~MeV for the $s$ and $d$ states, respectively.
To obtain $\Psi_\alpha^{(+)}$, $\psi_{pn}^i$ is
calculated up to $r_{pn}=100.0$~fm.

In the calculation of $\psi_{p \rm B}^j$ in the final channel,
we adopt a Woods-Saxon central potential as $U_{p{\rm B}}^{(\beta)}$ with radial
parameter $R_0 = 1.25\times 8^{1/3}~{\rm fm}$ and
diffuseness parameter $a_0 = 0.65~{\rm fm}$.
Its depth is determined to reproduce the proton separation
energy of $1.30$~MeV in the $p$ state.
The interaction between a point charge
and a uniformly charged sphere with the charge radius
2.5~fm is used as $V_{\rm C}$, which is used also in the
CDCC calculation in the initial channel.
The pseudostate method is also used for the final channel.
For the expansion of $\psi_{p \rm B}^j$ we take 20 Gaussian basis
functions with the minimum (maximum) range parameter of 1.0 (20.0)~fm.
We take into account the $s$, $p$, $d$, $f$, and $g$ waves of
$\psi_{p \rm B}^j$ with maximum values of $\varepsilon_{p{\rm B}}^j$
of 70, 75, 85, 90, and 70~MeV, respectively.
$\psi_{p \rm B}^j$ is calculated up to $r_{p{\rm B}}=100.0$~fm.

For $U_{p \rm B}^{(\alpha)}$, $U_{n \rm B}^{(\alpha)}$, and
$U_{n \rm B}^{(\beta)}$,
we adopt the nucleon global optical potential for $p$-shell
nuclei by
Watson {\it et al.}~\cite{Watson} (WA).
The non local correction
proposed by Timofeyuk and Johnson~\cite{nlTJ1,nlTJ2,nlTJ3} (TJ)
to the nucleon distorting potentials of the initial channel
is used. The calculated energy shift~\cite{nlTJ1,nlTJ2,nlTJ3} with
the above-mentioned $p$-$n$ model is 17.8~MeV in the c.m. frame.
We thus evaluate $U_{p \rm B}^{(\alpha)}$ and $U_{n \rm B}^{(\alpha)}$
at 33.0~MeV in the laboratory frame, which is shifted from the incident energy of 14.4~MeV/nucleon.
The non local correction to $U_{n\rm B}^{(\beta)}$
is made following Perey and Buck~\cite{PereyBuck} with the non-local parameter
$\beta=0.85$~fm.

For describing the transfer reaction, Eq.~(\ref{Tmat1}) is integrated
over $\boldsymbol{r}_{\alpha}$ and $\boldsymbol{r}_{\beta}$
up to 25.0 and 20.0~fm, respectively.
The number of partial waves for $\chi_\alpha^{ii_0(+)}$ and
$\chi_\beta^{jj_0(-)}$ is $25$.
As mentioned above, we include only the $s$ states of $\psi_{pn}^i$,
consisting of the ground and discretized-continuum states,
in the calculation of the $T$ matrix of the transfer process.
It should be noted that the coupling between the $s$ and $d$ states
of $\psi_{pn}^i$ is taken into account in the calculation of $\Psi_{\alpha}^{(+)}$
with the CDCC method.
It is found that $D_{pn}^i$ defined below by Eq.~\eqref{Dpn} is negligibly small for the $d$ states
of the deuteron, which justifies their neglect in the transfer
process.

\subsection{Asymptotic normalization coefficient (ANC) and astrophysical
factor $S_{18}(0)$}
\label{result2}
%
\begin{figure}[t]
\begin{center}
\includegraphics[width=0.48\textwidth,clip]{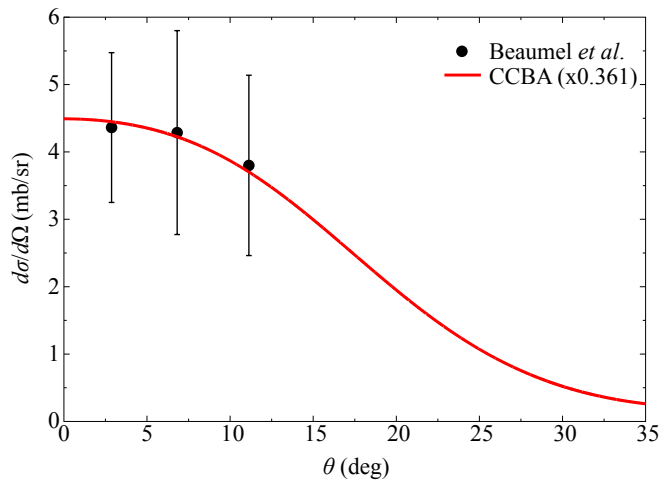}
\caption{(Color online)
 Cross section of the transfer reaction $^{8}$B($d$,$n$)$^{9}$C at
 14.4~MeV/nucleon as a function of the neutron emission angle in
 the c.m. frame.
 The result of the CCBA (solid line) is
 normalized to the experimental data~\cite{Beaumel}.
}
\label{fig2}
\end{center}
\end{figure}
We show in Fig.~\ref{fig2} the cross section of the transfer reaction $^{8}$B($d$,$n$)$^{9}$C at
14.4~MeV/nucleon as a function of the neutron emission angle in the c.m.
frame. The solid
line shows the CCBA result.
We have normalized 
the result
to reproduce the experimental data~\cite{Beaumel} multiplied by ${\cal S}=0.361$.
Note that, from the present transfer reaction, ${\cal S}$ cannot be determined
because the reaction is peripheral, as will be confirmed below.
Instead, the asymptotic normalization
coefficient (ANC)~\cite{ANC,sbat,breakupS18}
$C_{p {}^{8}\mbox{\scriptsize B}}^{^{9}\mbox{\scriptsize C}}$
for the overlap of the $^9$C wave function with the $p$-$^8$B(g.s.)
configuration is well determined.
From ${\cal S}$ and the so-called single-particle ANC of $\psi_{p \rm B}^{j_0}$,
one can obtain
the ANC: $(C_{p {}^{8}\mbox{\scriptsize B}}^{^{9}\mbox{\scriptsize C}})^2=0.59$~fm$^{-1}$.

Accuracy of the value of the ANC depends on how the transfer reaction
$^{8}$B($d$,$n$)$^{9}$C
is peripheral with respect to $r_{p \rm B}$. This can be examined
by estimating the dependence of
$C_{p {}^{8}\mbox{\scriptsize B}}^{^{9}\mbox{\scriptsize C}}$
on the parameters of $U_{p\rm{B}}^{(\beta)}$; each of $R_0$ and $a_0$
is changed by 20\%.
As mentioned above, we put a constraint on the depth of the potential
so that the proton separation energy is reproduced.
It is found that, by this change of $R_0$ and $a_0$, 
$(C_{p {}^{8}\mbox{\scriptsize B}}^{^{9}\mbox{\scriptsize C}})^2$
varies by only 2\%, which indicates the  peripherality of the
transfer reaction and guarantees the reliability of
$C_{p {}^{8}\mbox{\scriptsize B}}^{^{9}\mbox{\scriptsize C}}$.

Uncertainty due to the distorting potential is estimated by using another
nucleon global potential set for $p$-shell nuclei.
We adopt the parameter set by Dave and Gould~\cite{Dave-Gould} (DG).
Since the incident energy corrected with the TJ prescription for
nonlocality, 33.0~MeV, is out of the range of the DG parametrization,
we see the difference between the values of ANC calculated with WA and DG
potentials, both without the nonlocal correction. As a result,
the uncertainty of the ANC coming from the optical potential is found
to be 3\%.

By compiling the uncertainties due to
peripherality (2\%) and the optical potential (3\%)
as well as
the experimental error of 22\%~\cite{Beaumel}, we obtain
$(C_{p {}^{8}\mbox{\scriptsize B}}^{^{9}\mbox{\scriptsize C}})^2
=0.59 \pm 0.02~{\rm (theor.)} \pm 0.13~{\rm (exp.)}$~fm$^{-1}$,
where (theor.) and (exp.), respectively, stand for the theoretical and
experimental uncertainties.
Using the proportionality of
$(C_{p {}^{8}\mbox{\scriptsize B}}^{^{9}\mbox{\scriptsize C}})^2$
to $S_{18}(0)$, we have
\beq
S_{18}(0)
=22 \pm 1~{\rm (theor.)} \pm 5~{\rm (exp.)}~{\rm eVb}.
\eeq

\subsection{Breakup effects of $d$ and $^9$C on transfer cross section}
\label{result3}
%
\begin{figure}[b]
\begin{center}
\includegraphics[width=0.48\textwidth,clip]{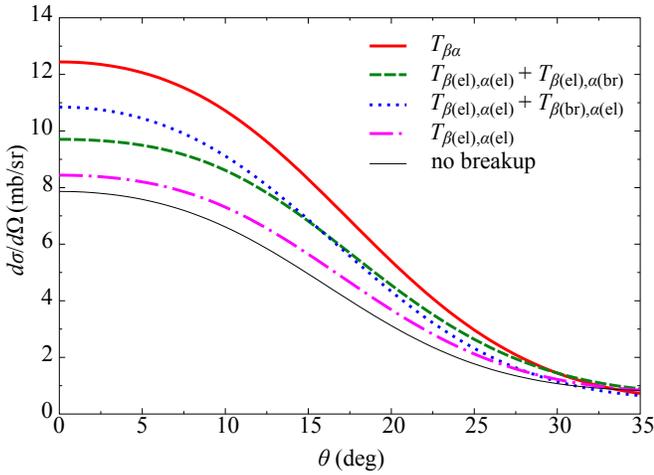}
\caption{(Color online)
 Breakup effects of $d$ and $^9$C on the cross
 section of $^{8}$B($d$,$n$)$^{9}$C at 14.4~MeV/nucleon.
 The thick solid and thin solid lines show, respectively, the results with and without
 the breakup states of both $d$ and $^9$C.
 The dashed (dotted) line represents the result with the
 breakup states of $d$ ($^9$C) in the transition matrix $T_{\beta\alpha}$ being neglected.
The cross section corresponding to the ET is shown by the dash-dotted line.
See the text for detail.
}
\label{fig3}
\end{center}
\end{figure}
The result for $S_{18}(0)$ in the present study,
$22 \pm 6~{\rm eV~b}$,
is somewhat smaller than the result from the previous analysis ($45\pm 13$~eV~b)
extracted from the same experimental data~\cite{Beaumel} with the
distorted-wave Born approximation (DWBA), which does not
explicitly take into account the breakup states of nuclei.
In this section we discuss this difference in view of the
breakup effects of $d$ and $^9$C in the transfer reaction.
In Fig.~\ref{fig3}, we show by the thick (thin) solid line
the cross section of $^{8}$B($d$,$n$)$^{9}$C at 14.4~MeV/nucleon
calculated with (without) the breakup states of both $d$ and $^9$C.
Inclusion of the breakup channels gives a large increase of about 58\% in
the cross section at 0$^\circ$.

To see this in more detail, we decompose the $T$ matrix into
\begin{align}
T_{\beta \alpha}
&=
 T_{\beta({\rm el}),\alpha({\rm el})}
+T_{\beta({\rm el}),\alpha({\rm br})}
\nonumber \\
&+T_{\beta({\rm br}),\alpha({\rm el})}
 +T_{\beta({\rm br}),\alpha({\rm br})}
,\label{Tdiv}
\end{align}
\begin{align}
T_{\beta({\rm el}),\alpha({\rm el})}
&\equiv
\left\langle
\psi_{p{\rm B}}^{j_0}
\chi_\beta^{j_0j_0(-)}
\Big|V_{pn}\Big|
\psi_{pn}^{i_0}
\chi_\alpha^{i_0i_0(+)}
\right\rangle
,\label{Telel}
\\
T_{\beta({\rm el}),\alpha({\rm br})}
&\equiv
\left\langle
\psi_{p{\rm B}}^{j_0}
\chi_\beta^{j_0j_0(-)}
\Big|V_{pn}\Big|
\sum_{i \ne i_0}
\psi_{pn}^{i}
\chi_\alpha^{ii_0(+)}
\right\rangle
,\label{Telbr}
\\
T_{\beta({\rm br}),\alpha({\rm el})}
&\equiv
\left\langle
\sum_{j \ne j_0}
\psi_{p{\rm B}}^{j}
\chi_\beta^{jj_0(-)}
\Big|V_{pn}\Big|
\psi_{pn}^{i_0}
\chi_\alpha^{i_0i_0(+)}
\right\rangle
,\label{Tbrel}
\\
T_{\beta({\rm br}),\alpha({\rm br})}
&\equiv
\left\langle
\sum_{j \ne j_0}
\psi_{p{\rm B}}^{j}
\chi_\beta^{jj_0(-)}
\Big|V_{pn}\Big|
\sum_{i \ne i_0}
\psi_{pn}^{i}
\chi_\alpha^{ii_0(+)}
\right\rangle
.\label{Tbrbr}
\end{align}
The $T$ matrix with the subscript $\gamma{\rm (el)}$ and
$\gamma{\rm (br)}$ corresponds
to the elastic transfer (ET) and the breakup transfer (BT) in the $\gamma$ channel, respectively.
The dash-dotted line in Fig.~\ref{fig3} shows the cross section due to
the ET described by $T_{\beta{\rm (el)},\alpha{\rm (el)}}$.
Note that $T_{\beta{\rm (el)},\alpha{\rm (el)}}$ includes the breakup
effects as the back-coupling between the elastic channel and the breakup
channels for both $d$ and $^9$C.
However, the small difference between the thin solid line and
the dash-dotted line indicates that those back-coupling effects
are not significant in the present case.
The dashed line shows the result including the breakup states of only $d$,
which is about 23\% larger than that shown by the thin solid line at $0^\circ$.
It is also found that the transfer cross section through the breakup states
of $d$ is less than 1\% of that shown by the dashed line. We thus conclude that
the increase in the cross section caused by the breakup states of $d$
is due to the interference between $T_{\beta{\rm (el)},\alpha{\rm (el)}}$
and $T_{\beta{\rm (el)},\alpha{\rm (br)}}$.
This conclusion holds also for the role of the breakup states of $^9$C;
large interference between
$T_{\beta{\rm (el)},\alpha{\rm (el)}}$ and $T_{\beta{\rm (br)},\alpha{\rm (el)}}$
increases the cross section by about 38\% at $0^\circ$, as shown by
the dotted line.
Furthermore, it is found numerically that the contribution of $T_{\beta{\rm
(br)},\alpha{\rm (br)}}$ to the cross section is negligibly small.

These properties of the numerical result can be understood as follows.
If we make the adiabatic approximation~\cite{AD1,AD2,TJ99} to
$\Psi_{\alpha}^{(+)}$, we have
\begin{align}
\Psi_{\alpha}^{(+)}(\boldsymbol{r}_{pn},\boldsymbol{r}_{\alpha})
&\approx
\psi_{pn}^{i_0}(\boldsymbol{r}_{pn})
\chi_\alpha^{\rm AD(+)}(\boldsymbol{r}_{pn},\boldsymbol{r}_{\alpha})
.\label{ADwf}
\end{align}
The adiabatic wave function $\chi_\alpha^{\rm AD(+)}$ satisfies
\begin{align}
\left[
K_{\boldsymbol{r}_{\alpha}}
\!\!+U_{p \rm B}^{(\alpha)} ({r}_{p \rm B})
\!+\!U_{n \rm B}^{(\alpha)} ({r}_{n \rm B})
\!- E_\alpha
\right]
\!\!
\chi_\alpha^{\rm AD(+)}(\boldsymbol{r}_{pn},\boldsymbol{r}_{\alpha})
\!=0
,\label{ScheqAD}
\end{align}
where $E_\alpha = E + \varepsilon^{i_0}_{pn}$. The
$\boldsymbol{r}_{pn}$ dependence of $U_{N \rm B}^{(\alpha)}$ ($N=p$ or $n$)
gives that of $\chi_\alpha^{\rm AD(+)}$.
Consequently, $\Psi_{\alpha}^{(+)}$ contains
not only the elastic-channel but also the breakup-channel components:
\beq
\chi_\alpha^{ii_0{\rm AD}(+)}(\boldsymbol{r}_{\alpha})
\equiv
\langle
\psi_{pn}^{i}(\boldsymbol{r}_{pn})
|
\psi_{pn}^{i_0}(\boldsymbol{r}_{pn})
\chi_\alpha^{\rm AD(+)}(\boldsymbol{r}_{pn},\boldsymbol{r}_{\alpha})
\rangle.
\eeq
The $\boldsymbol{r}_{pn}$ dependence of $U_{N \rm B}^{(\alpha)}$ is,
however, quite weak within the range of $V_{pn}$. Then one can expect
that, for $\chi_\alpha^{ii_0{\rm AD}(+)}$ with $i\neq i_0$,
the amplitude would be much smaller than that of
$\chi_\alpha^{i_0i_0{\rm AD}(+)}$ and
the phase would be very similar to that of $\chi_\alpha^{i_0i_0{\rm AD}(+)}$.
The former is the reason for the very small contribution of the
BT and the latter is that for the constructive
interference between the ET and BT amplitudes.
These properties have been confirmed numerically.
This interpretation of the breakup effects can also be applied to
$\Psi_\beta^{(-)}$ in the final channel.
It should be noted that the adiabatic approximation~\cite{AD1,AD2,TJ99}
itself is found to work well;
it makes $C_{p{}^{8}\mbox{\scriptsize B}}^{^{9}\mbox{\scriptsize C}}$ smaller
by about 6\% (12\%) when applied to $\Psi_\alpha^{(+)}$ $\left(\Psi_\beta^{(-)}\right)$.

%
\begin{figure}[b]
\begin{center}
\includegraphics[width=0.48\textwidth,clip]{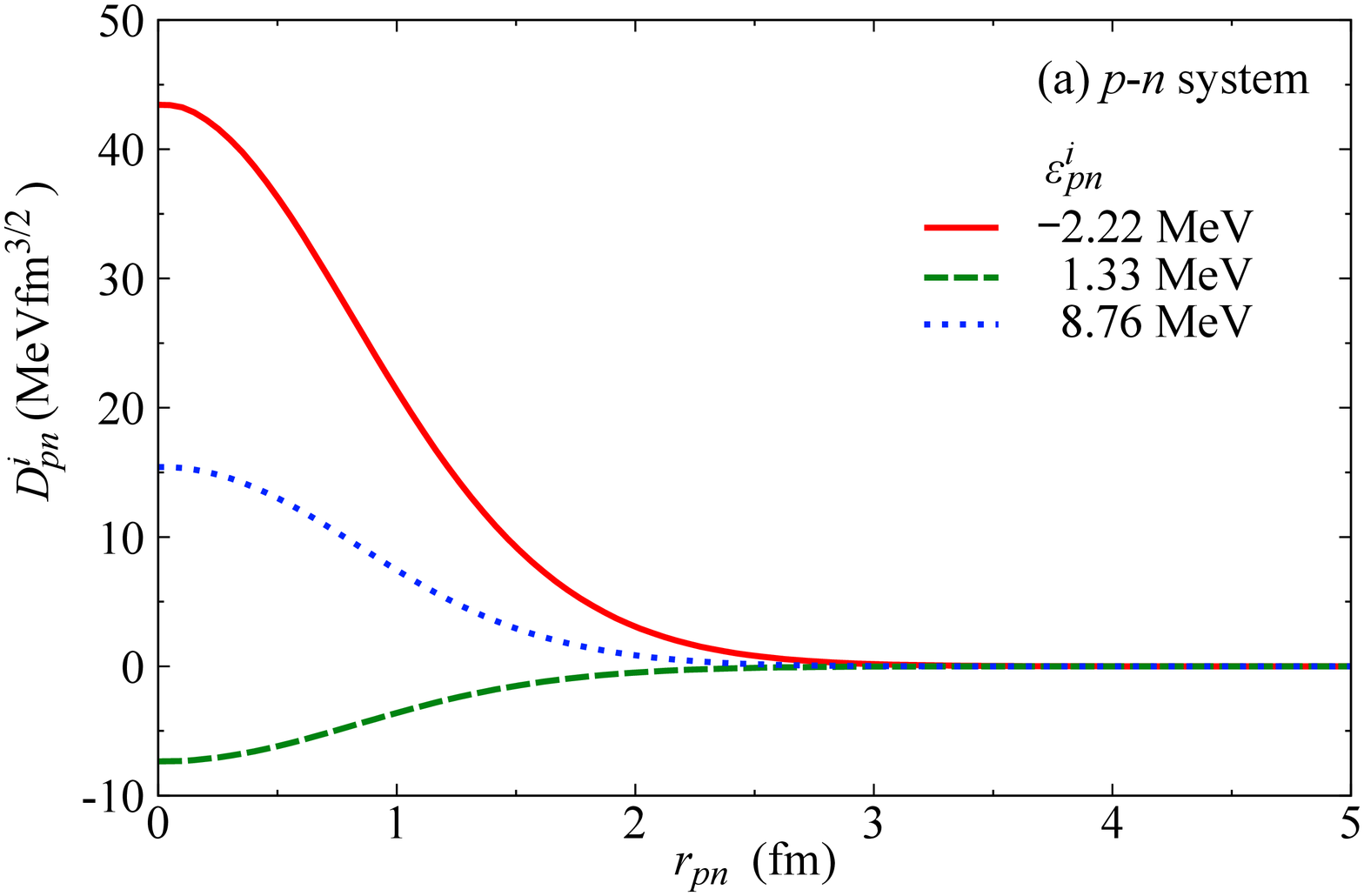}
\includegraphics[width=0.48\textwidth,clip]{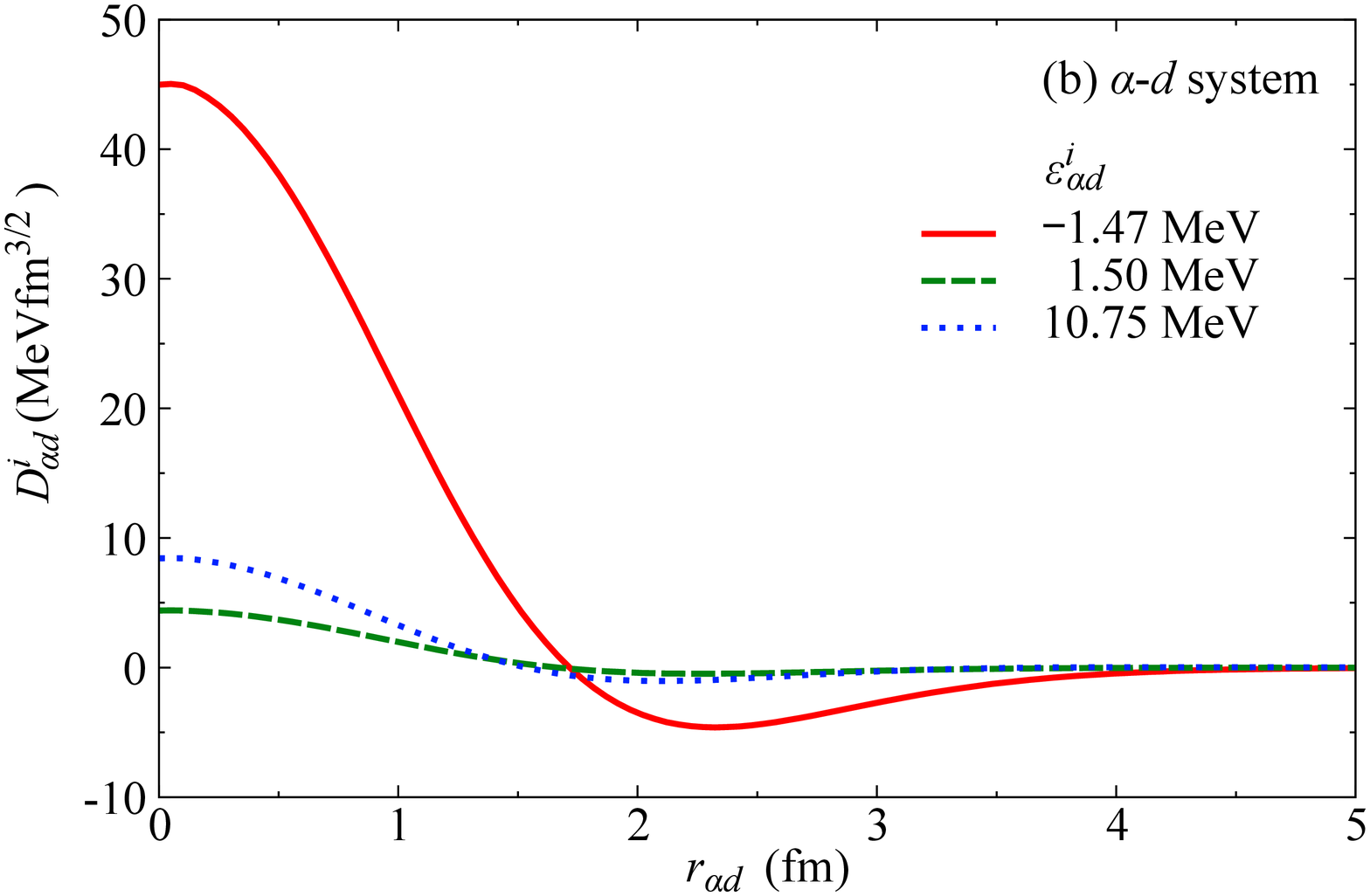}
\caption{(Color online)
 (a) $D_{pn}^i$ for several $i$th states with the eigenenergy
 $\varepsilon_{pn}^i$.
 (b) Same as in panel (a) but for the $\alpha$-$d$ system.
}
\label{fig4}
\end{center}
\end{figure}
%

As mentioned above, the back-coupling effects are found to be small
in the present case.
In fact, if we evaluate $C_{p {}^{8}\mbox{\scriptsize B}}^{^{9}\mbox{\scriptsize C}}$
and $S_{18}(0)$ from the thin solid line, we obtain
$(C_{p {}^{8}\mbox{\scriptsize B}}^{^{9}\mbox{\scriptsize C}})^2=0.95$~fm$^{-1}$
and $S_{18}(0)=36~{\rm eV~b}$.
This value is, within only about 2\% difference, consistent with
the result corresponding to the D1-N1 set for the distorting potentials,
$(C_{p {}^{8}\mbox{\scriptsize B}}^{^{9}\mbox{\scriptsize
C}})^2=0.97$~fm$^{-1}$, shown in Table~1 of Ref.~\cite{Beaumel};
N1 corresponds to the WA potential.
We have confirmed by our DWBA calculation that the result with the
D1-N1 set agrees well with
the thin solid line in Fig.~\ref{fig3}.
From these findings we conclude that inclusion of the breakup
states of both $d$ and $^9$C is necessary to accurately describe
the transfer reaction, which gives quite a large
increase in the cross section, that is, decrease in $S_{18}(0)$.

The non-negligible BT component in each channel is
opposite to what was found in
the analysis~\cite{sbat} of $^{13}$C($^6$Li,$d$)$^{17}$O
below the Coulomb barrier energy, in which breakup effects
of $^6$Li ($=\alpha+d$) were investigated.
Below we discuss the difference between the breakup
properties of $d$ and $^6$Li
in the two reactions.
The origin of the difference can be understood from
the behavior of $D_{ pn}^i$ defined by
\begin{align}
D_{ pn}^i({r}_{ pn})
&=
V_{ pn}(r_{ pn})
\phi_{pn}^i(r_{ pn})
,\label{Dpn}
\end{align}
where $\phi_{pn}^i$ is the radial part of $\psi_{pn}^i$.
We show in Fig.~\ref{fig4}(a) $D_{pn}^i$
for some $s$-wave eigenstates of $d$;
the eigenvalue $\varepsilon_{pn}^i$ is given in the legend.
Similarly, we plot in Fig.~\ref{fig4}(b)
$D_{\alpha d}^i(r_{\alpha d})=V_{\alpha d}(r_{\alpha d})
\phi_{\alpha d}^i(r_{\alpha d})$ for the $\alpha$-$d$ system;
the two-range Gaussian interaction $V_{\alpha d}$ given
in Ref.~\cite{Sakuragi} is adopted to generate the radial part
$\phi_{\alpha d}^i$
of the $s$-wave eigenstate $\psi_{\alpha d}^i$.

In Figs.~\ref{fig4}(a) and (b), respectively,
$D_{pn}^i$ and $D_{\alpha d}^i$ for some eigenstates are plotted.
One sees that the amplitude of $D_{pn}^i$ for breakup states
(the dashed and dotted lines) are comparable to that of
$D_{pn}^{i_0}$ (solid line).
On the other hand, $D_{\alpha d}^i$ for the breakup states
are much smaller than $D_{\alpha d}^{i_0}$, which is found
to be due to the Coulomb interaction between $\alpha$ and $d$.
Thus, the difference in the BT components between
the $^{8}$B($d$,$n$)$^{9}$C and $^{13}$C($^6$Li,$d$)$^{17}$O
reactions can be understood.
It should be noted that a large value of $D^i$ for a breakup state
does not necessarily give a large BT cross section,
because even in this case $\chi_\alpha^{i i_0}$ can be small
as a result of the channel couplings. Furthermore, the importance
of the back-coupling effect depends on the reaction system
in a non trivial manner.

\subsection{Formalism of finite-range correction for CCBA transition amplitude
and finite-range effect on transfer cross section}
\label{result4}
In this section we describe a procedure for an FRC
to the ZR CCBA transition matrix. The essence of this correction is
similar to that given in Ref.~\cite{SatchDNR}, except that the present
method is based on a three-body reaction model
including continuum states of both the projectile and the
residual nucleus.
The integral expression of Eq.~(\ref{Tmat1}), with Eq.~(\ref{CDCCf}), is given by
\begin{align}
T_{\beta \alpha}
=&
\sum_j
\int d\boldsymbol{r}_{pn} d\boldsymbol{r}_{\alpha}
\chi_{\beta}^{jj_0(-)*}(\boldsymbol{r}_{\beta})
\psi_{p \rm B}^{j*}(\boldsymbol{r}_{p \rm B})
V_{pn}(r_{pn})
\nonumber \\
& \times
\Psi_{\alpha}^{(+)}(\boldsymbol{r}_{pn},\boldsymbol{r}_{\alpha})
.\label{Tmat2}
\end{align}
By using
\begin{align}
\psi_{p \rm B}^{j*}(\boldsymbol{r}_{p{\rm B}})
&=
\psi_{p \rm B}^{j*}(\boldsymbol{r}_{\alpha} +\sigma \boldsymbol{r}_{ pn} )
=
e^{\sigma \nabla_{\boldsymbol{r}_{p \rm B}} \cdot \boldsymbol{r}_{ pn}}
\psi_{p \rm B}^{j*}(\boldsymbol{r}_{\alpha})
,\nonumber \\
\chi_{\beta}^{jj_0(-)*}(\boldsymbol{r}_{\beta})
&=
\chi_{\beta}^{jj_0(-)*}(\tau^{-1} \boldsymbol{r}_{\alpha} +\xi \boldsymbol{r}_{ pn} )
\nonumber \\
&=
e^{\tau \xi \nabla_{\boldsymbol{r}_\beta} \cdot \boldsymbol{r}_{ pn}}
\chi_{\beta}^{jj_0(-)*}(\tau^{-1} \boldsymbol{r}_{\alpha})
\label{DW}
\end{align}
with $\sigma=1/2$, $\tau=9/8$, and $\xi=\sigma/\tau -1$,
Eq. (\ref{Tmat2}) can be rewritten as
\begin{align}
T_{\beta\alpha}
=
&
\sum_j
\int d\boldsymbol{r}_{ pn} d\boldsymbol{r}_{\alpha}
e^{(\sigma \nabla_{\boldsymbol{r}_{p \rm B}} +\tau\xi \nabla_{\boldsymbol{r}_\beta}) \cdot \boldsymbol{r}_{ pn}}
\nonumber
\\
&\times
\chi_{\beta}^{jj_0(-)*}(\tau^{-1} \boldsymbol{r}_{\alpha})
\psi_{ p \rm B}^{j*}(\boldsymbol{r}_{\alpha})
V_{ pn}(r_{ pn})
\Psi_{\alpha}^{(+)}(\boldsymbol{r}_{pn},\boldsymbol{r}_{\alpha})
.\label{Tmat3}
\end{align}
It should be noted that
$\nabla_{\boldsymbol{r}_{p \rm B}}$ and $\nabla_{\boldsymbol{r}_\beta}$
operate on only $\psi_{p \rm B}^{j*}$ and $\chi_\beta^{jj_0(-)*}$, respectively.

As in Ref.~\cite{SatchDNR}, we use
\begin{align}
e^{(\sigma \nabla_{\boldsymbol{r}_{p \rm B}} +\tau\xi \nabla_{\boldsymbol{r}_\beta}) \cdot \boldsymbol{r}_{ pn}}
&\approx
1+\frac{1}{6}
(\sigma \nabla_{\boldsymbol{r}_{p \rm B}} +\tau\xi \nabla_{\boldsymbol{r}_\beta})^2
r_{ pn}^2
.\label{expexp}
\end{align}
Here, we assume that only the $s$-wave component of the deuteron wave
function contributes to the $T$ matrix, which has eliminated the
first-order term of the expansion series in Eq.~(\ref{expexp});
justification of this assumption is given in Sec.~\ref{result1}.
With the local energy approximation~\cite{SatchDNR}, one may find
\begin{align}
T_{\beta\alpha}
\approx &
\sum_{j}
\int d\boldsymbol{r}_{ pn} d\boldsymbol{r}_{\alpha}
\chi_{\beta}^{jj_0(-)*}(\tau^{-1} \boldsymbol{r}_{\alpha})
\psi_{ p \rm B}^{j*}(\boldsymbol{r}_{\alpha})
V_{ pn}(r_{ pn}) \nonumber \\
&
\times \hat{F}_{\rm LEA}
\Psi_{\alpha}^{(+)}(\boldsymbol{r}_{pn},\boldsymbol{r}_{\alpha})
\label{Tmat4}
\end{align}
with
\begin{align}
\hat{F}_{\rm LEA}\equiv
1+\frac{1}{6} r_{ pn}^2
\frac{2\mu_{pn}}{\hbar^2}
&\Big[
 U^{(\beta)}_{p \rm B}(r_{p \rm B})
+U^{(\beta)}_{n \rm B}(r_{n \rm B}) +\Delta V_{\rm C}
\nonumber \\
&
-U_{p \rm B}^{(\alpha)}(r_{p \rm B})
-U_{n \rm B}^{(\alpha)}(r_{n \rm B}) -h_{pn}
\Big]
\label{FLEAop}
\end{align}
and
\beq
\Delta V_{\rm C}\equiv
V_{\rm C}(r_{p{\rm B}}) - V_{\rm C}(r_\alpha),
\label{delvc}
\eeq
where $\mu_{pn}$ is the reduced mass of the $p$-$n$ system.
Here we assume $\Delta V_{\rm C}\sim 0$.
Note that, if we include the Coulomb breakup in the initial channel,
$V_{\rm C}(r_\alpha)$ is replaced with $V_{\rm C}(r_{p{\rm B}})$,
which results in $\Delta V_{\rm C}=0$.
Using $\boldsymbol{r}_{p \rm B}=\boldsymbol{r}_{\alpha} +\sigma \boldsymbol{r}_{pn}$
and $\boldsymbol{r}_{n \rm B}=\boldsymbol{r}_{\alpha} -\sigma \boldsymbol{r}_{pn}$,
we make the following expansion:
\begin{align}
U_{p \rm B}^{(\gamma)}(r_{p \rm B})
&\approx
U_{p \rm B}^{(\gamma)}(r_{\alpha})
+\left[
\nabla_{\boldsymbol{r}_{\alpha}} U_{p \rm B}^{(\gamma)}(r_{\alpha})
\right]
\cdot \sigma \boldsymbol{r}_{pn}
,\label{UpB}
\\
U_{n \rm B}^{(\gamma)}(r_{n \rm B})
&\approx
U_{n \rm B}^{(\gamma)}(r_{\alpha})
-\left[
\nabla_{\boldsymbol{r}_{\alpha}} U_{n \rm B}^{(\gamma)}(r_{\alpha})
\right]
\cdot \sigma \boldsymbol{r}_{pn}
.\label{UnB}
\end{align}
The second terms of Eqs.~(\ref{UpB}) and~(\ref{UnB}) vanish
after being integrated over $\boldsymbol{r}_{pn}$,
because we consider only the $s$-wave states of $\psi_{pn}^i$,
as mentioned above.
By using Eqs.~(\ref{CDCCi}) and (\ref{CDwf}), we then obtain
\begin{align}
T_{\beta\alpha}
\approx
\sum_{ij}&
\int d\boldsymbol{r}_{\alpha}
\chi_{\beta}^{jj_0(-)*}(\tau^{-1} \boldsymbol{r}_{\alpha})
\psi_{ p \rm B}^{j*}(\boldsymbol{r}_{\alpha})
\nonumber \\
&\times
D_0^i
F_{\rm LEA}^i(r_{\alpha})
\chi_{\alpha}^{ii_0(+)}(\boldsymbol{r}_{\alpha})
\label{Tmat6}
\end{align}
with
\begin{align}
F_{\rm LEA}^i(r_{\alpha})
\equiv
1+\frac{\rho_i^2}{6}
\frac{2\mu_{pn}}{\hbar^2}
&\Big[
 U^{(\beta)}_{p \rm B}(r_\alpha)
+U^{(\beta)}_{n \rm B}(r_\alpha)
\nonumber \\
&
-U_{p \rm B}^{(\alpha)}(r_\alpha)
-U_{n \rm B}^{(\alpha)}(r_\alpha) -\varepsilon_{pn}^i
\Big]
.
\label{FLEA}
\end{align}
In Eqs.~(\ref{Tmat6}) and (\ref{FLEA}) $D_0^i$ and $\rho^2_i$ are defined by
\begin{align}
D_0^i
&=\sqrt{4\pi}
\int d r_{ pn}
r^2_{ pn} D_{ pn}^i({r}_{ pn})
,\label{D0}
\end{align}
\begin{align}
\rho^2_i
&=
\frac{\displaystyle\int d r_{ pn}
r_{ pn}^4 D_{ pn}^i({r}_{ pn})}
{\displaystyle\int d r_{ pn}
r^2_{ pn} D_{ pn}^i({r}_{ pn})}
.\label{rho}
\end{align}
Thus, the integration over ${\boldsymbol{r}}_{ pn}$ is factored out in
the evaluation of the $T$ matrix.
It should be noted that the FRC function $F_{\rm LEA}^i$ does not depend on $j$.

If we take only the first term on the right-hand-side (r.h.s) of Eq.~(\ref{FLEA}),
we obtain a $T$ matrix with the ZR approximation to $D_{ pn}^i$:
\begin{align}
D_{ pn}^i({r}_{ pn})
=
\frac{D_0^i}{\sqrt{4\pi}}
\delta(r_{ pn})
\label{ZR}.
\end{align}
Therefore, the second term on the r.h.s. of Eq.~(\ref{FLEA}) is
regarded as the FRC to the ZR calculation.
Equations~(\ref{Tmat6}) and (\ref{FLEA}) give a natural extension of the
FRC proposed in Ref.~\cite{SatchDNR} that can be used in the CCBA formalism.

When the breakup states in the final channel are neglected
as in the previous study~\cite{transferS17},
Eq.~(\ref{FLEA}) reduces to
\begin{align}
F_{\rm LEA}^i(r_{\alpha})
=
1+\frac{\rho_i^2}{6}
\frac{2\mu_{pn}}{\hbar^2}
&\Big[
 U^{(\beta)}_{p \rm B}(r_\alpha)
+U^{(\beta)}(\tau^{-1}r_\alpha)
\nonumber \\
&
-U_{p \rm B}^{(\alpha)}(r_\alpha)
-U_{n \rm B}^{(\alpha)}(r_\alpha) -\varepsilon_{pn}^i
\Big]
,
\label{FLEA2}
\end{align}
where $U^{(\beta)}$ is the distorting potential for the $n$-$^9$C
scattering wave function.
This expression is useful when we adopt the CDCC wave function in only the initial channel.

Further simplification of Eq.~(\ref{FLEA}) can be done if
$U_{p \rm B}^{(\beta)}\approx U^{(\alpha)}_{p \rm B}$
and $U_{n \rm B}^{(\beta)} \approx
U_{n \rm B}^{(\alpha)}$, that is,
\begin{align}
F_{\rm LEA}^i(r_\alpha)
&\approx
1-\frac{\rho_i^2}{6}
\frac{2\mu_{ pn}}{\hbar^2}
\varepsilon_{pn}^i
.\label{FLEA2}
\end{align}
By definition, $\varepsilon_{pn}^i$ is negative for the ground state ($i=i_0$)
and positive for the breakup states ($i \neq i_0$).
Thus, we can see from Eq.~(\ref{FLEA2}) that for the transfer process through
the deuteron ground state, the ET, the FRC
increases the $T$-matrix element. On the other hand, for the transfer process
through the breakup states of $d$, the BT, the correction
gives a decrease in the $T$-matrix element. This behavior is useful to interpret
the difference between the results of the ZR and FR calculations,
as shown below.
It should be noted that $\rho_i^2$ can be negative when $\varepsilon_{pn}^i$ is very
large. However, the contribution of such state to the $T$ matrix is found
to be negligibly small. Note also that in the actual calculation
we use Eq.~(\ref{FLEA}); Eq.~(\ref{FLEA2}) is used
just for interpretation of the numerical result.

%
\begin{figure}[htpb]
\begin{center}
\includegraphics[width=0.48\textwidth,clip]{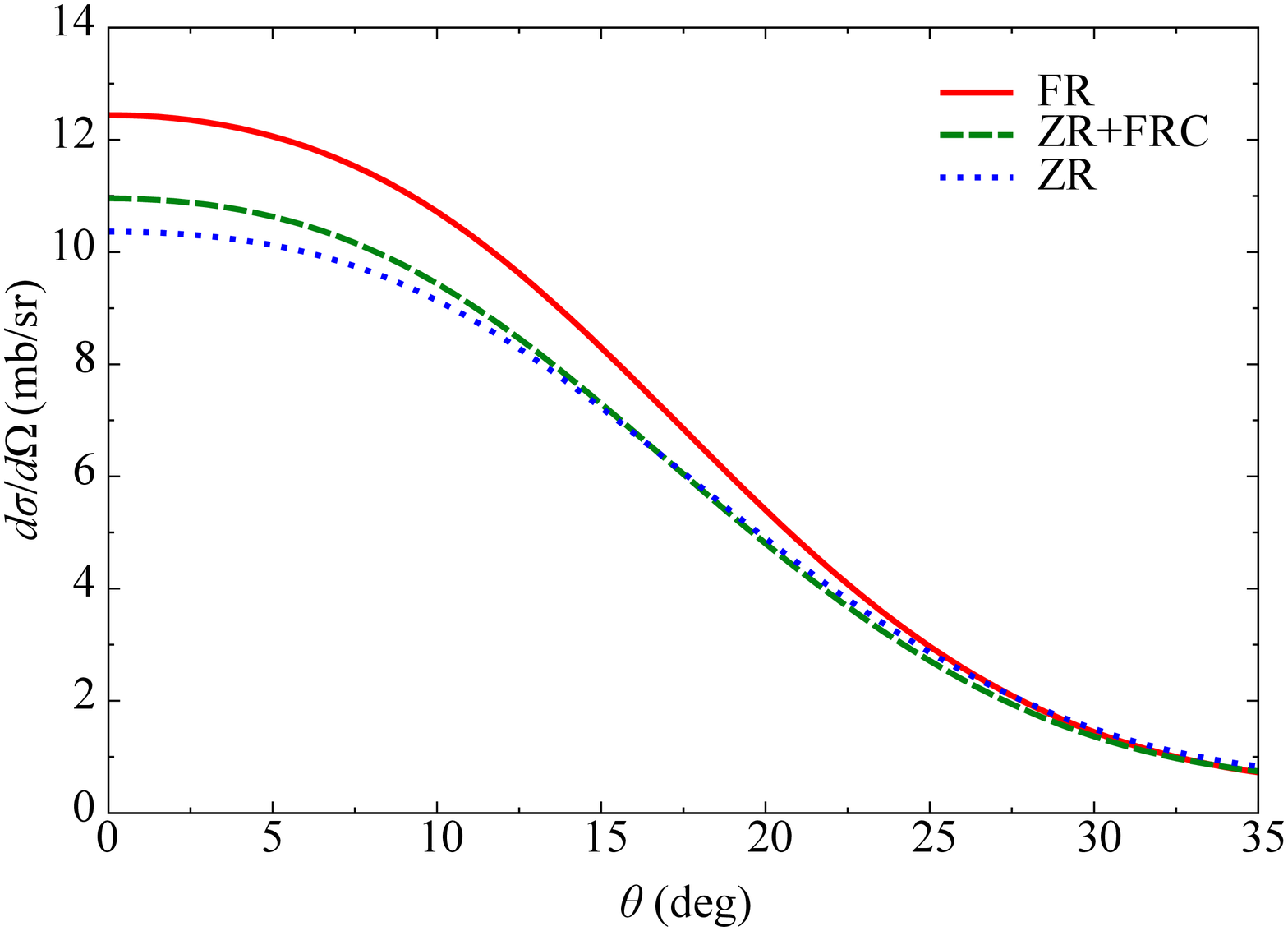}
\caption{(Color online)
 CCBA results of the FR calculation (solid line),
 the ZR calculation (dotted line), and the ZR calculation
 with the FRC (dashed line).
}
\label{fig5}
\end{center}
\end{figure}
%

%
\begin{figure}[htpb]
\begin{center}
\includegraphics[width=0.48\textwidth,clip]{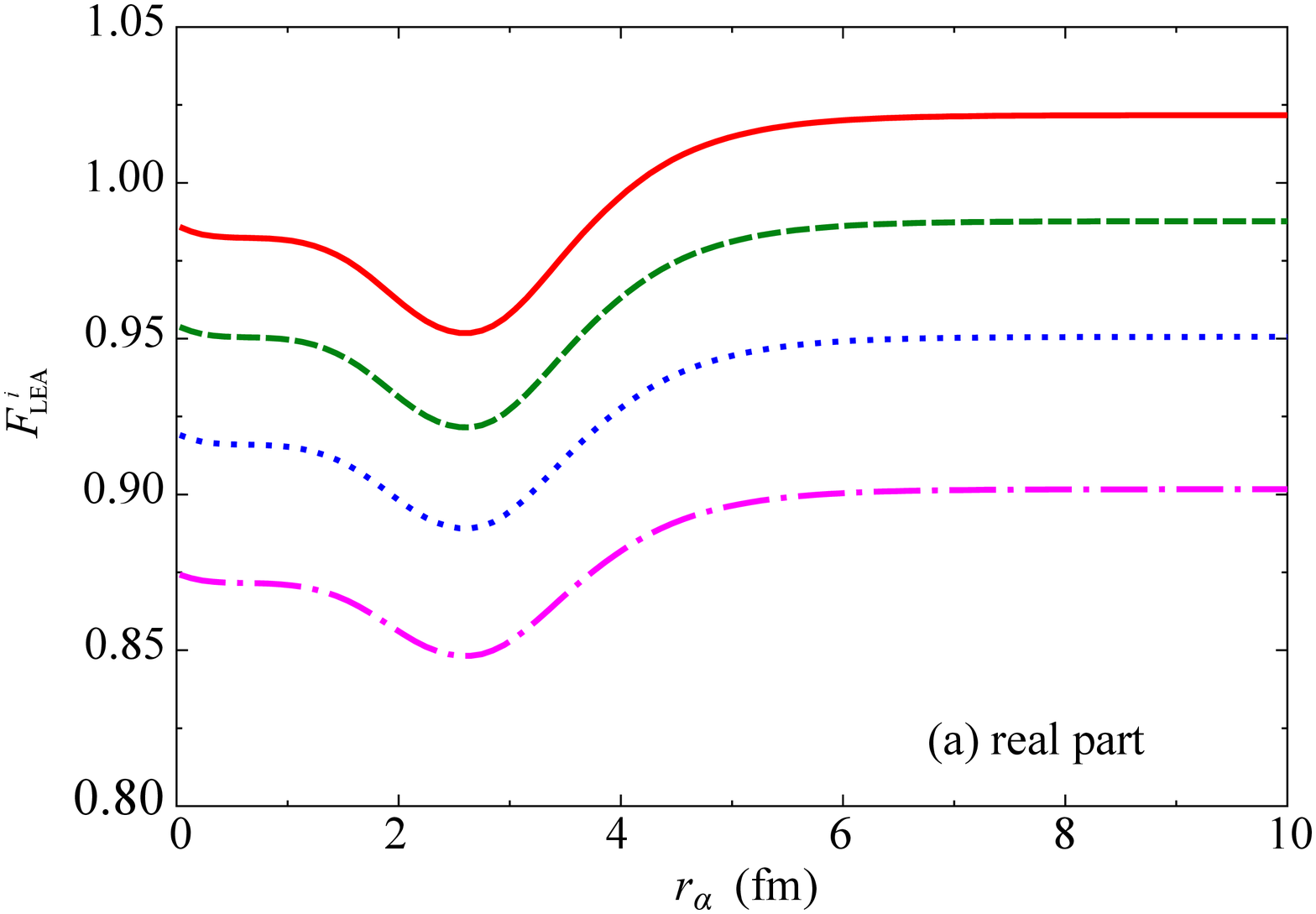}
\includegraphics[width=0.48\textwidth,clip]{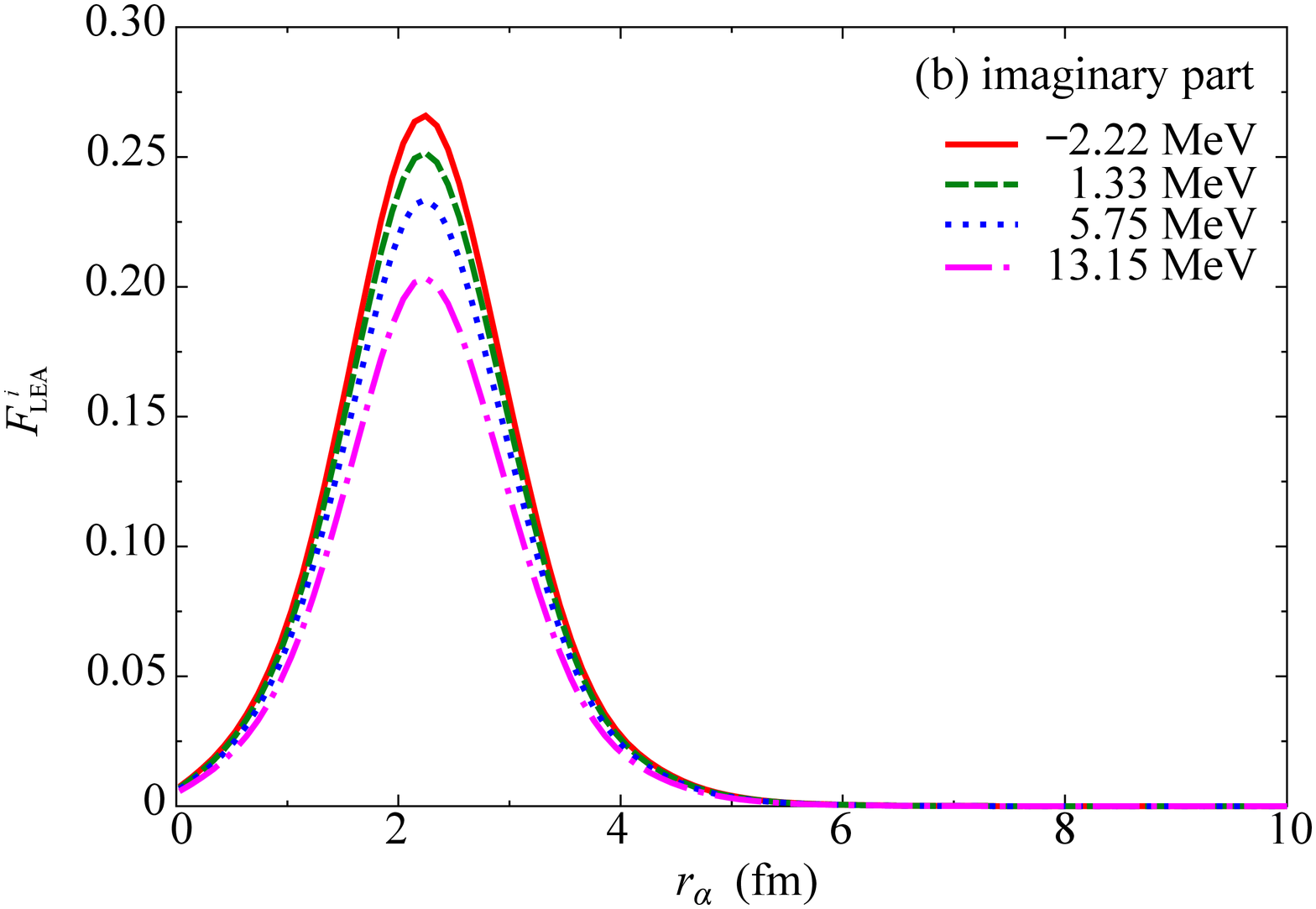}
\caption{(Color online)
 (a) Real and (b) imaginary parts
 of the correction function $F_{\rm LEA}^i$ defined by Eq.~(\ref{FLEA}).
 Each line corresponds to the result with $\varepsilon_{pn}^i$
 specified in the legends.
}
\label{fig6}
\end{center}
\end{figure}
We show in
Fig.~\ref{fig5} the results obtained by the FR
calculation (solid line), the ZR calculation (dotted line),
and the ZR calculation with the FRC described
by Eqs.~(\ref{Tmat6}) and (\ref{FLEA})
(dashed line).
One finds that the FR effect gives about a 20\% increase in the cross section
at $\theta=0^\circ$.
The FRC works well qualitatively but is not sufficient to get good
agreement with the solid line.
This suggests that the FR effect found in $^{8}$B($d$,$n$)$^{9}$C at
14.4~MeV/nucleon 
contains a higher-order component that cannot be included in the
present procedure.

The correction function
$F^i_{\rm LEA}$ of Eq.~(\ref{FLEA}) is plotted in Fig.~\ref{fig6};
panels (a) and (b) correspond to the real and imaginary parts of
$F^i_{\rm LEA}$, respectively.
It is found that $F^i_{\rm LEA}$ has a nontrivial
behavior in the interior region, say, $r_\alpha \la 6$~fm.
As clarified in Sec.~\ref{result2}, however,
the $^{8}$B($d$,$n$)$^{9}$C reaction at 14.4~MeV/nucleon
is peripheral with respect
to $r_{p \rm B}$, which is the same as $r_\alpha$ in the ZR limit.
Thus, the contribution of $F^i_{\rm LEA}$ in the interior
region to the $T$ matrix is expected to be very small.
In this case, a simple estimation of the FR effect based on Eq.~(\ref{FLEA2})
works well. At higher incident energies, where we have less peripherality,
the FR effect can change significantly.

\section{Summary}
\label{summary}
%
\begin{figure}[b]
\begin{center}
\includegraphics[width=0.48\textwidth,clip]{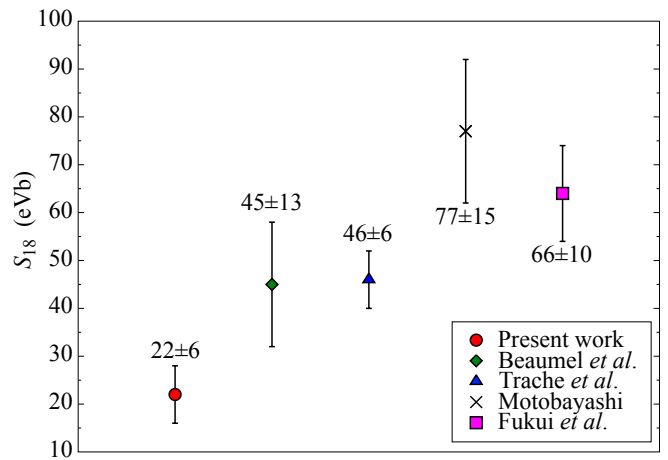}
\caption{(Color online)
 $S_{18}(0)$ in the present work (circle) is
 compared with the results evaluated from the $^{8}$B($d$,$n$)$^{9}$C reaction
 (diamond)~\cite{Beaumel} and values extracted from $^9$C breakup
 reactions (triangle~\cite{Trache}, cross~\cite{Motobayashi}, and square~\cite{breakupS18}).
}
\label{fig7}
\end{center}
\end{figure}
We have analyzed
the transfer reaction $^{8}$B($d$,$n$)$^{9}$C at 14.4~MeV/nucleon
by means of the $p+n+{}^{8}$B three-body coupled-channels framework.
The ANC of $^9$C in the $p$-$^8$B(g.s.) configuration,
$C_{p{}^{8}\mbox{\scriptsize B}}^{^{9}\mbox{\scriptsize C}}$,
and the astrophysical factor at zero energy, $S_{18}(0)$,
for the $^8$B($p$,$\gamma$)$^9$C reaction have been determined.
Our results are $(C_{p{}^{8}\mbox{\scriptsize B}}^{^{9}\mbox{\scriptsize C}})^2
=0.59 \pm 0.15$~fm$^{-1}$
and $S_{18}(0) =22 \pm 6~{\rm eVb}$.
It is found that the breakup states of both $d$ and $^9$C increase the transfer
cross section through the interference between the ET
and BT amplitudes. As a result, the present result is
smaller than the previous value~\cite{Beaumel} extracted from the same
experimental data by about 51\%.
The back-coupling effects on the elastic channel are found to be small.

We proposed a new prescription of the FRC to the
ZR calculation of the $T$ matrix, which can be used in the CCBA
formalism. For the $^{8}$B($d$,$n$)$^{9}$C reaction at 14.4~MeV/nucleon,
the FRC is not sufficient to reproduce the result of the
FR calculation, indicating the importance of higher-order correction terms.
The FR effect on the transfer reaction considered turns out
to be about 20\%.

In Fig.~\ref{fig7} we compare the present result for $S_{18}(0)$ with
previous results extracted from indirect measurements.
As mentioned, we obtained a smaller $S_{18}(0)$
than that of Ref.~\cite{Beaumel} because of the contribution of
$d$ and $^9$C breakup states.
The present result is not consistent with
the result of a three-body model analysis~\cite{breakupS18}
of the inclusive~\cite{Trache} and exclusive~\cite{Motobayashi} $^9$C
breakup reactions within $2\sigma$.
Further investigation is necessary to understand the reason for this
discrepancy. Extension of the present framework to include
breakup channels of $^8$B as well as the three-body model
description of $^9$C will be important future work.
Another possible reason for the discrepancy in $S_{18}(0)$
is the Pauli blocking effect on the transfer reaction~\cite{Paulieff1,Paulieff2}.
Antisymmetrization between a nucleon in $d$ and that in $^8$B in
calculation of the $d$-$^8$B three-body wave function will be an important
subject.

In Ref.~\cite{S18mirror}, $S_{18}(0)=44 \pm 11$~eV~b was
extracted from $^8$Li($d$,$p$)$^9$Li, which is the mirror reaction to
$^{8}$B($d$,$n$)$^{9}$C, by means of the DWBA.
It will be interesting to estimate breakup effects of $d$ in this
mirror reaction.
Furthermore, a compilation of the ANCs for the $p$-shell nuclei has been
made recently~\cite{STA}, in which
$C_{p{}^{8}\mbox{\scriptsize B}}^{^{9}\mbox{\scriptsize C}}=1.080~{\rm fm}^{-1}$ was reported.
It will be important to elucidate the difference between this value and the present
result.

\begin{acknowledgments}
The authors thank Y.~Iseri, Y.~Kanada-En'yo, and K.~Minomo for helpful discussions.
This research was supported in part by a Grant-in-Aid of the Japan
Society for the Promotion of Science (JSPS).
\end{acknowledgments}

\nocite{*}




\begin{thebibliography}{99}

\bibitem{Wiescher}
M. Wiescher, J. G\"{o}rres, S. Graff, L. Buchman, and F.-K. Thieleman,
Astrophys. J. \textbf{343}, 352 (1989).

\bibitem{Trache}
L. Trache, F. Carstoiu, A. M. Mukhamedzhanov, and R. E. Tribble,
Phys. Rev. C \textbf{66}, 035801 (2002).

\bibitem{Motobayashi}
T. Motobayashi, Nucl. Phys. \textbf{A718}, 101c (2003).

\bibitem{Beaumel}
D. Beaumel \textit{et al.}, Phys. Lett. \textbf{B514}, 226 (2001).

\bibitem{breakupS18}
T. Fukui, K. Ogata, K. Minomo, and M. Yahiro, Phys. Rev. C \textbf{86}, 022801(R) (2012).

\bibitem{transferS17}
K. Ogata, M. Yahiro, Y. Iseri, and M. Kamimura,
Phys. Rev. C \textbf{67}, 011602(R) (2003).

\bibitem{CDCC1}
M. Kamimura {\it et al.}, 
Prog. Theor. Phys. Suppl. No.~89, 1 (1986).

\bibitem{CDCC2}
N.~Austern {\it et al.}, 
Phys. Rep. \textbf{154}, 125 (1987).

\bibitem{CDCC3}
M.~Yahiro {\it et al.}, 
Prog. Theor. Exp. Phys. \textbf{2012}, 01A206 (2012).

\bibitem{Aus89}
N.~Austern, M.~Yahiro, and M.~Kawai,
Phys. Rev. Lett. {\bf 63}, 2649(1989).

\bibitem{Aus96}
N.~Austern, M.~Kawai, and M.~Yahiro,
Phys. Rev. C {\bf 53}, 314 (1996).

\bibitem{sbat}
T. Fukui, K. Ogata, and M. Yahiro,
Prog. Theor. Phys.  {\bf 125}, 1193 (2011).



\bibitem{Ohmura}
T.~Ohmura {\it et al.},
Prog. Theor. Phys. \textbf{43}, 347 (1970).

\bibitem{Mat03}
T.~Matsumoto, T.~Kamizato, K.~Ogata, Y.~Iseri, E.~Hiyama,
M.~Kamimura, and M.~Yahiro,
Phys. Rev. C {\bf 68}, 064607 (2003).

\bibitem{Watson}
B.~A.~Watson, P.~P.~Singh, and R.~E.~Segel,
Phys. Rev. \textbf{182}, 997 (1969).

\bibitem{nlTJ1}
N.~K.~Timofeyuk and R.~C.~Johnson,
Phys. Rev. Lett. \textbf{110}, 112501 (2013).

\bibitem{nlTJ2}
N.~K.~Timofeyuk and R.~C.~Johnson,
Phys. Rev. C \textbf{87}, 064610 (2013).

\bibitem{nlTJ3}
R.~C.~Johnson and N.~K.~Timofeyuk,
Phys. Rev. C \textbf{89}, 024605 (2014).

\bibitem{PereyBuck}
G. Perey and B. Buck,
Nucl. Phys. \textbf{32}, 353 (1962).

\bibitem{ANC}
A. M. Mukhamedzhanov and N. K. Timofeyuk,
Yad. Fiz. {\bf 51}, 679 (1990)
[Sov. J. Nucl. Phys. {\bf 51}, 431 (1990)].

\bibitem{Dave-Gould}
J.~H.~Dave and C.~R.~Gould,
Phys. Rev. C \textbf{28}, 2212 (1983).

\bibitem{AD1}
H.~Amakawa, S.~Yamaji, A.~Mori, and K.~Yazaki,
Phys. Lett. \textbf{B82}, 13 (1979).

\bibitem{AD2}
M.~A.~ Nagarajan, I.~J.~Thompson, and R.~C.~Johnson
Nucl. Phys. \textbf{A385}, 525 (1982).


\bibitem{TJ99}
N.~K.~Timofeyuk and R.~C.~Johnson,
Phys. Rev. C \textbf{59}, 1545 (1999).

\bibitem{Sakuragi}
Y.~Sakuragi, M.~Yahiro, and M.~Kamimura,
Prog. Theor. Phys. Suppl. \textbf{89}, 136 (1986).

\bibitem{SatchDNR}
G. R. Satchler, {\it Direct Nuclear Reactions} (Oxford University Press, New York, 1983), p. 245.

\bibitem{Paulieff1}
W.~S.~Pong and  N.~Austern,
Ann. Phys. (NY) \textbf{93}, 369 (1975).

\bibitem{Paulieff2}
R.~C.~Johnson, N.~Austern, and M.~H.~Lopes,
Phys. Rev. C \textbf{26}, 348 (1982).

\bibitem{S18mirror}
B. Guo {\it et al.}, 
Nucl. Phys. \textbf{A761}, 162 (2005).

\bibitem{STA}
N.~K.~Timofeyuk,
Phys. Rev. C \textbf{88}, 044315 (2013), and references therein.


\end{thebibliography}
\end{document}